\documentstyle[12pt]{article}

\def\be{\begin{equation}}
\def\ee{\end{equation}}
\def\ben{\begin{displaymath}}
\def\een{\end{displaymath}}
\def\ba{\begin{array}{c}}
\def\bal{\begin{array}{l}}
\def\ea{\end{array}}
\def\p{\partial}
\begin{document}


\vspace{1.5cm}

 \begin{center}{\Large \bf
A return to observability near exceptional points in a schematic
${\cal PT}-$symmetric model
  }\end{center}

\vspace{10mm}

 \begin{center}

 {\bf Miloslav Znojil}

 \vspace{3mm}
\'{U}stav jadern\'e fyziky AV \v{C}R,

250 68 \v{R}e\v{z},

Czech Republic

{e-mail: znojil@ujf.cas.cz}

\vspace{3mm}

\vspace{5mm}


\end{center}

\vspace{5mm}

\section*{Abstract}

Many indefinite-metric (often called pseudo-Hermitian or ${\cal
PT}-$symmetric) quantum models $H$ prove ``physical" (i.e.,
Hermitian with respect to an innovated, {\em ad hoc} scalar
product) inside a characteristic domain of parameters ${\cal D}$.
This means that the energies get complex (= unobservable) beyond
the boundary $\p {\cal D}$ (= Kato's ``exceptional points", EPs).
In a solvable example we detect an enlargement of ${\cal D}$
caused by the emergence of a new degree of freedom. We conjecture
that such a beneficial mechanism of a return to the real spectrum
near EPs may be generic and largely model-independent.

\newpage

\section{Introduction and summary}

Over virtually any model in quantum phenomenology one initially
feels urged to consider {\em all} the relevant degrees of freedom
in the corresponding Langangian ${\cal L}$ or Hamiltonian ${\cal
H}$. This tendency is limited by the imperatives of the
tractability of calculations and of a feasibility of making
measurable predictions. Thus, for example, for an electron moving
in a very strong external Coulomb field, an exhaustive theoretical
analysis requires the full-fledged formalism of relativistic
quantum field theory but some of the measurable properties of the
bound states are still very satisfactorily predicted by the mere
quantum-mechanical, exactly solvable Dirac-equation model
\cite{Greiner}.

One of the most characteristic features of many ``reduced" models
of the latter type is that their reliability (i.e., in an abstract
formulation, the negligibility of relevance of their ``frozen"
degrees of freedom) may vary with some of their dynamical
parameters. Thus, in the same illustration one reveals that when
the external field becomes strong enough, some of the
Dirac-field-excitation components of the system enter the scene
and become directly coupled to the motion of the electron itself
\cite{Greiner}. In such a dynamical regime the Dirac-equation
predictions fail and, formally and typically, the energies of the
electron itself become complex.

On a simplified model-building level the similar ``paradoxes" in
the behavior of the energies may be explained using the
parity-pseudo-Hermitian Hamiltonians $H$. They are often called
${\cal PT}-$symmetric,  with the defining property $H \neq
H^\dagger= {\cal T}\,H\,{\cal T}= {\cal P}\,H\,{\cal P}$ and with
a formal operator-conjugation ${\cal T}$ mimicking the time
reversal and with ${\cal P}$ representing the parity (cf. also
Appendix A for more details).

The latter Hamiltonians still can be re-interpreted as
self-adjoint (i.e., observable) but one must restrict their set of
dynamical parameters (i.e., of couplings etc) to a certain
subdomain ${\cal D}^{(physical)}$ on which their spectrum remains
real. In the context of Quantum Mechanics, more details may be
found in the review paper \cite{Geyer}, while an immediate and
inspiring extension of such a recipe to Field Theory has only been
proposed much more recently, in ref.~\cite{BM}.

In our recent letter \cite{Hendrik} we introduced, for
illustrative purposes, a two-state model with the generic
one-parametric ${\cal PT}-$symmetric Hamiltonian
  \be
 H^{(2)} = \left (
 \begin{array}{cc}
 -1&a\\
 -a&1
 \ea
 \right )\,,
 \ \ \ \ \ \
  \ \ \ \
 {\cal P}^{(2)} = \left (
 \begin{array}{cc}
 1&0\\
 0&-1
 \ea
 \right )\,.
 \label{pseudoh}
  \ee
We also explained there the existence of a nontrivial formal
relationship between our simple model (\ref{pseudoh}) and its more
standard (and phenomenologically ambitious) differential-operator
predecessors or analogues $H^{({\cal PT})}$ (say, of
refs.~\cite{others}). In essence, this relationship is based on
the replacement of $H^{({\cal PT})}$ by their equivalent
infinite-dimensional matrix representants $H^{(\infty)}$ (in a
suitable basis) and, subsequently, by their variational,
$N-$dimensional truncated-matrix approximants $H^{(N)}$ with $N <
\infty$. In such a context it still makes sense to coin the name
``parities" for the corresponding ``indefinite-metric" matrices
${\cal P}^{(N)}$ which enter the finite-dimensional
pseudo-Hermiticity property $H^\dagger\,{\cal P}^{(N)}={\cal
P}^{(N)}H$ of the matrix toy Hamiltonians $H=H^{(N)}$.

The key purpose of our present short paper is to show that the
simple matrix models of the form (\ref{pseudoh}) can say a lot
about the interpretation of the general ${\cal PT}-$symmetric
Hamiltonians in the critical regime where their energies are about
to complexify. In this sense we intend to complement now our
remark \cite{Hendrik} on the schematic model $H^{(2)}$ by a few
new and interesting observations based on a tentative immersion of
the two-dimensional system in a generic three-dimensional one,
 \be
 H^{(3)}=
 \left (\begin {array}{cc|c}
 -1&a&0
 \\
  -a&1&b\\
  \hline
  0&-b&3+c
  \end {array}\right )\,\ \ \ \ \
 \ \ \ \ \ \ \ \ \ \
 {\cal P}^{(3)}=
 \left (\begin {array}{cc|c} 1&0&0\\
 0&-1&0
 \\
 \hline
  0&0&1
  \end {array}\right )\,.
 \label{sekoust}
 \ee
Here, the standard Jacobi rotation has been employed in setting
zeros in the corners of $H^{(3)}$ so that the three-parametric
example (\ref{sekoust}) preserves the full generality of its
one-parametric predecessor (\ref{pseudoh}).

In a continuation of the study \cite{Hendrik} we shall be able to
show that and how a fairly satisfactory insight in several
qualitative though, up to now, not too well understood properties
of the general ${\cal PT}-$symmetric models can be deduced from
the mere comparison of the virtually elementary models
(\ref{pseudoh}) and (\ref{sekoust}). First of all, such a
comparison will enable us to study some of the aspects of the
above-mentioned conflict between the use of the models with ``too
few" and ``too many" degrees of freedom by choosing the models
$H^{(2)}$ and $H^{(3)}$ as their respective representatives. In
section \ref{kolik} we emphasize that the vanishing of one of the
two coupling constants in the ``universal-like" model $H^{(3)}$
leads directly to the sample ``reduced" model of the form
$H^{(2)}$. A few relevant results on $H^{(2)}$ of
ref.~\cite{Hendrik} are summarized there for the sake of
completeness as well.

Our mathematical encouragement lies in the exact non-numerical
tractability of the ``universal" model $H^{(3)}$. In section
\ref{core} the feasibility of quantitative calculations will
enable us to extend the results of ref.~\cite{Hendrik} to the
richer model $H^{(3)}$ where we set $c=0$ for the sake of
simplicity. In particular, we shall show that the formula for the
boundary $\p {\cal D}(H^{(3)})$ of the domain where all the
energies remain real can be written in closed form.

A core of our message will be formulated in
section~\ref{generalka} where we show how the growth of ${\cal
D}(H^{(N)})$ from $N=2$ to $N=3$ shifts the Kato's exceptional
points \cite{Kato}. In subsection \ref{subsef} we emphasize that
in the closest vicinity of such an exceptional point at $N=2$,
even the weakest coupling to an ``observer channel" induces a
steady growth of the modified quasi-Hermiticity domain ${\cal
D}(H^{(2)})$. In subsection \ref{tripll}, this observation is
extended to the manifestly non-perturbative regime with the
strongest couplings near the doubly exceptional points where all
the three energy levels coincide.

One could summarize our present message as opening the possibility
of a systematic amendment of various ${\cal PT}-$symmetric models
near exceptional points via a re-activation of certain ``frozen"
degrees of freedom. In this sense, more work is still needed to
confirm that our qualitative observations might stay valid far
beyond the range of the present study.

Let us add that in Appendix A we complemented our discussion by a
concise review of literature showing the physical background and,
perhaps, broader relevance and possible impact of our schematic
models. In a more technical remark of Appendix B we finally show
that the role of $c\neq 0$ in our model $H^{(3)}$ can be
rightfully ignored as not too essential.

\section{Simulated changes of degrees of freedom\label{kolik}}

\subsection{Decoupling an observer state: $b \to
0$ and  $H^{(3)} \to H^{(2)}$
  \label{dvakratdva}  }

Once we start from the illustrative example (\ref{pseudoh}),
letter \cite{Hendrik} tells us that

\begin{itemize}

 \item
the eigenenergies  remain real and non-degenerate whenever $a^2<
1$, $$ E_\pm = \pm \sqrt{1-a^2}$$ so that we may set $ a = \cos
\alpha$ with $ \alpha \in (0,\pi)$ in ${\cal D}(H^{(2)})$;
 \item
the necessary {\it ad hoc} scalar products of Appendix A are
obtainable via a metric operator in (\ref{newpr}). The choice of
this operator is ambiguous, with its elements numbered by an
overall multiplicative constant and by another real parameter
$\gamma \in [0, \pi/2)$,
 \be
 \Theta \sim
 \left (
 \begin{array}{cc}
 1+\xi&-\cos \alpha\\
 -\cos \alpha&1-\xi
 \ea
 \right )\,,\ \ \ \ \ \
 \xi = \sin \alpha \sin \gamma\,;
  \label{finalre}
 \ee
 \item
at {\em both} the exceptional points $\alpha^{(EP)} = 0, \pi$ of
the boundary $\p {\cal D}$, {\em all} the matrices
$\Theta=\Theta(\gamma)$ cease to be invertible so that
$\Theta^{-1}$ [needed in definition (\ref{newconj}) below] ceases
to exist.

\end{itemize}

 \noindent
One may note that at the EP singularities the geometric and
algebraic multiplicities of eigenvalues become different. Some
energies complexify immediately beyond these points. Near the
points of the boundary $\p {\cal D}$, all the predictions of
quantum mechanics may be more sensitive to perturbations and must
be examined particularly carefully.

\subsection{A re-activated degree of freedom:
$b \neq 0$ and  $H^{(2)} \to H^{(3)}$ \label{trikrattri}}

In the language of physics, one should contemplate introducing
some new degree(s) of freedom near every exceptional point. The
majority of the current Hamiltonians $H$ [say, of the
differential-operator form (\ref{ptsp}) discussed in Appendix A]
does not offer a feasible option of this type. In contrast, the
finite-dimensional matrix models can incorporate a new degree of
freedom very easily, via an elementary increase of their
dimension.

In an illustration let us first recollect that in the
two-dimensional model of preceding paragraph we have $a^{(EP)}=\pm
1$. In the vicinity of these exceptional points we may set
$a=\pm(1-\varepsilon)$ with a real and sufficiently small
$\varepsilon$ which remains positive inside ${\cal D}(H^{(2)})$,
vanishes in the EP regime and gets negative outside the domain.

An increase of the dimension $N$ in $H^{(N)}$ from 2 to 3 should
be accompanied by a coupling of the submatrix $H^{(2)}$ to a new,
``observer" element of the basis. The resulting ${\cal
PT}-$symmetric three-state matrix model (\ref{sekoust}) contains a
new real coupling $b$ and another real parameter $c \neq -2, -4 $.
Note that the presence of the two vanishing elements in $H^{(3)}$
does not weaken its generality since the corresponding two-by-two
submatrix remains Hermitian and is assumed pre-diagonalized.

\section{Exceptional points in the model $H^{(3)}$
\label{core} }

A pairwise attraction of the energy levels mediated by the
variations of the couplings $a$ and $b$ in $H^{(3)}$ should
control the changes of the spectrum in full analogy with the
generic two-state model. The role of the third parameter $c$ is
less essential and the discussion of its influence is postponed to
Appendix B. Now we set $c=0$ and insert our toy Hamiltonian
(\ref{sekoust}) in the three-state Schr\"{o}dinger equation. Its
determinantal secular equation for energies
 \be
 -{{\it E}}^{3}+3\,{{\it E}}^{2}+\left (-{a}^{2}+1-{b}^{2}\right
 ){ \it E}-3+3\,{a}^{2}-{b}^{2}=0
 \label{sekouspro}
 \ee
is solvable in closed form, via the well known Cardano formulae.
Cardano formulae offer the roots of eq.~(\ref{sekouspro}) in the
compact and non-numerical form which is, unfortunately, not too
suitable for the specification of the domain ${\cal D}$. In a
preparatory step, let us analyze a few simpler special cases of
eq.~(\ref{sekouspro}), therefore.

\subsection{Boundary
$\p {\cal D}$  at $a=c=0$ or $b=c=0$ \label{hladka}}

A quick inspection of eq. (\ref{sekouspro}) reveals that the
cheapest information about $\p {\cal D}$  becomes available when
$ab=0$.  We choose $b=0$ and decouple $H^{(3)}=H^{(2)}\bigoplus
H^{(1)}$. The analysis degenerates to the two-dimensional problem
and restricts the admissible values of $a$ to the following open
interval,
 \ben
   a \in \left . {\cal D}\right |_{\,b=0}=(-1,1).
 \een
This means that at $b=0$ the ``observer" level $E_2=3$ stays
decoupled and it does not vary with $a$ at all, while the two
other levels (i.e., $E_0=-1$ and $E_1=1$ at $a=0$) become
attracted in proportion to the strength $a\neq 0$ of the
non-Hermiticity.

A completely analogous situation is encountered at $a=0$. In this
case it is comfortable to shift $E \to E- 2$ and get another
section of quasi-Hermiticity domain in closed form,
 \ben
  b \in \left . {\cal D}\right |_{\,a=0}=(-1,1).
 \een
The genuine three-state phenomena may only occur when {\em both}
$a$ and $b$ remain non-zero, making all the three energy levels
{\em mutually} attracted.

\subsection{The regime of simultaneous attraction, $a \neq 0 \neq b$
  \label{tasmhle} }



The set of the exceptional points forms the boundary
$\partial{\cal D}$ which connects the above-mentioned four
exceptional points in the $a - b$ plane. Its shape (see Figure 1)
may be deduced from secular eq.~(\ref{sekouspro}) by a
``brute-force" numerical technique. There also exists its
non-numerical description replacing the non-degenerate triplet of
energies $E$ by the doubly degenerate energy $z=1+\beta$ {\em
plus} a separate, ``observer" third energy value $y=-1+2\alpha$.

In the case of $b>a>0$ the value of $z$ should result from a
merger of $E_1=1$ with $E_2=3$ so that we may expect that $\beta
\in (0,1)$. Similarly, one may discuss the other orderings of $a$
and $b$. In parallel, the necessary universality of the attraction
of the levels (as observed above in the two-dimensional model)
implies that we must always have $\alpha > 0$. Thus, once we
replace eq. (\ref{sekouspro}) by its adapted polynomial EP version
of the same (third) degree in $E$,
 \be
 -(E-z)^2(E-y)=
 -{{\it E}}^{3}+(2z+y)\,{{\it E}}^{2}-\left (z^2+2yz\right
 ){ \it E}+yz^2
 =0\,
\label{sekousjo}
 \ee
the comparison of the quadratic terms in these two alternatives
gives us the constraint $\alpha+\beta=1$. Similarly, the
reparametrization of the linear and constant contributions leads
to the set of the two equations
 \ben
  a^2+b^2=4-3\beta^2, \ \ \ \ \ \
 3a^2-b^2=4-3\beta^2-2\beta^3
 \een
which may be re-read as the desired one-parametric definition of
the star-like shape of the curve forming the boundary
$\partial{\cal D}(H^{(3)})$, with $\beta \in (-1,1)$,
 \be
 a=a_\pm = \pm \sqrt{\frac{1}{2}\left (
 4-3\beta^2-\beta^3
 \right )}, \ \ \ \
 b=b_\pm = \pm \sqrt{\frac{1}{2}\left (
 4-3\beta^2+\beta^3
 \right )}\,.
 \label{paramo}
 \ee
We may notice that all the four above-mentioned special EP cases
are reproduced by this formula at $\beta=\pm 1$. Our analytic
description of the boundary $\p {\cal D}(H^{(3)})$ at $c=0$ is
complete.

\section{Beneficial effects of the growth of $b \neq 0$
 \label{generalka} }

Having the parametric definition (\ref{paramo}) of boundary $\p
{\cal D}(H^{(3)})$  at our disposal we know precisely where the
energy spectrum remains real. This observation has several
mathematically easy but physically appealing and relevant
consequences.

\subsection{A return of energies from complex to real
\label{subsef} }

We originally started from the two-level model $H^{(2)}$
containing a single parameter $a$. This means that in the language
of the ``complete" three-state model $H^{(3)}$ we worked in the
regime $b=0$ where the ``spectator" degree of freedom stayed
decoupled. Critical EP values were $a=a^{(EP)}=\pm 1$ so that the
energies lost their observability (i.e., the system collapsed) in
arbitrarily small vicinities of these EPs.

The situation changes when the real and, say, not too large
coupling $b \neq 0$ is switched on.

\subsection*{Lemma}

Whenever our two-level model $H^{(2)}$ becomes coupled to a
``spectator" state with $c=0$ and $b\neq 0$ in $H^{(3)}$, energies
remain real for $a\in (-1-\eta,1+\eta)$ at certain $\eta=\eta(b)>
0$.

\subsection*{Proof}

From the definition (\ref{paramo}) we may infer that, say, near
the EP where $(a,b)=(1,0)$ we may set $\beta = -1+ \varepsilon^2
 +{\cal O}(\varepsilon^3) $ and deduce that
 \ben
 b^{(EP)}=
 \sqrt{\frac{1}{2}\left [
 4-3(1-2\varepsilon^2)-(1-3\varepsilon^2)
 +{\cal O}(\varepsilon^3)
 \right ]}=
 \frac{3\varepsilon}{\sqrt{2}} +{\cal O}(\varepsilon^2).
 \een
In parallel we have
 \ben
  a^{(EP)}=
 \sqrt{\frac{1}{2}\left [
 4-3(1-2\varepsilon^2)+(1-3\varepsilon^2)
 +{\cal O}(\varepsilon^3)
 \right ]}=
 1+ \frac{3\varepsilon^2}{4}+{\cal O}(\varepsilon^3)
 \een
so that we come to the conclusion that the EP value of $a$ grows
with $|\,b|$,
 \ben
  a^{(EP)}=
 1+ \frac{\left [b^{(EP)}\right ]^2}{6}+{\cal O}
 \left \{ \left [b^{(EP)}\right ]^3 \right \}
 .
 \label{paramoto}
 \een
This means that we are allowed to choose a positive $\eta(b)={\cal
O}\left ( b^2\right )$.

{\bf QED.}

 \noindent
We see that whenever we introduce a new ``degree of freedom" by
setting $b \neq 0$, our system becomes stable in a non-empty
vicinity of any of the two original exceptional points $a=\pm 1$.
In the light of a ``generic" character of our example $H^{(3)}$,
one may expect similar behaviour of parametric dependence of the
reality of the spectrum in {\em all} the other (or at least
``many") ${\cal PT}-$symmetric models, irrespectively of their
particular matrix or differential-operator realization.

\subsection{Doubly exceptional character of
the strongest acceptable couplings
\label{tripll}}

Due to the mutual attraction of the energy levels in our
``generic" three-by-three example one may expect that there exist
certain ``doubly exceptional" points (DEPs) of the boundary $\p
{\cal D}(H^{(3)})$ where {\em all} the three energies coincide at
a triple root $E=z$ of the secular equation,
 \be
 (E-z)^3=0\,.
\label{sekousne}
 \ee
The comparison of the coefficients in eqs. (\ref{sekouspro}) and
(\ref{sekousne}) at $E^2$ gives $z=1$. The subsequent two
comparisons provide the two other coupled polynomial equations,
 \ben
 -3=1-{a}^{2}-{b}^{2}, \ \ \ \ \ \ 1 =-3+3\,{a}^{2}
 -{b}^{2}\,.
 \een
We get quickly $b^2=4-a^2$ from the first equation while the
assignment $a^2=2$ follows from the second one, in agreement with
eq.~(\ref{paramo}) at $\beta=0$.

We may conclude that in the light of Figure 1 and formulae
(\ref{paramo}), the boundary of the domain ${\cal D}(H^{(3)})$ of
the allowed real matrix elements $a$ and $b$ remains smooth not
only in the perturbative vicinity of the four points of subsection
\ref{hladka},
 $$(a,b) \in \{\, (1,0),
 \, (0,1), \,(-1,0), \,(0,-1) \,\},$$
but also at all the pairwise mergers of the real energies. The
spikes are encountered at the four ``maximal-coupling" vertices of
a circumscribed square,
 $$(a,b) \in \{\,
 (\sqrt{2},\sqrt{2}), \, (-\sqrt{2},\sqrt{2}),
 \,(-\sqrt{2},-\sqrt{2}), \,(\sqrt{2},-\sqrt{2}) \,\}$$
where one locates the DEP triple-energy mergers. The fourfold
symmetry of the whole boundary $\p {\cal D}^{(3)}$ of the
quasi-Hermiticity domain is just an accidental consequence of our
simplifying  choice of the vanishing spectral shift~$c=0$.

\section*{Acknowledgement}

Supported by GA\v{C}R, grant Nr. 202/07/1307.

\section*{Figure captions}

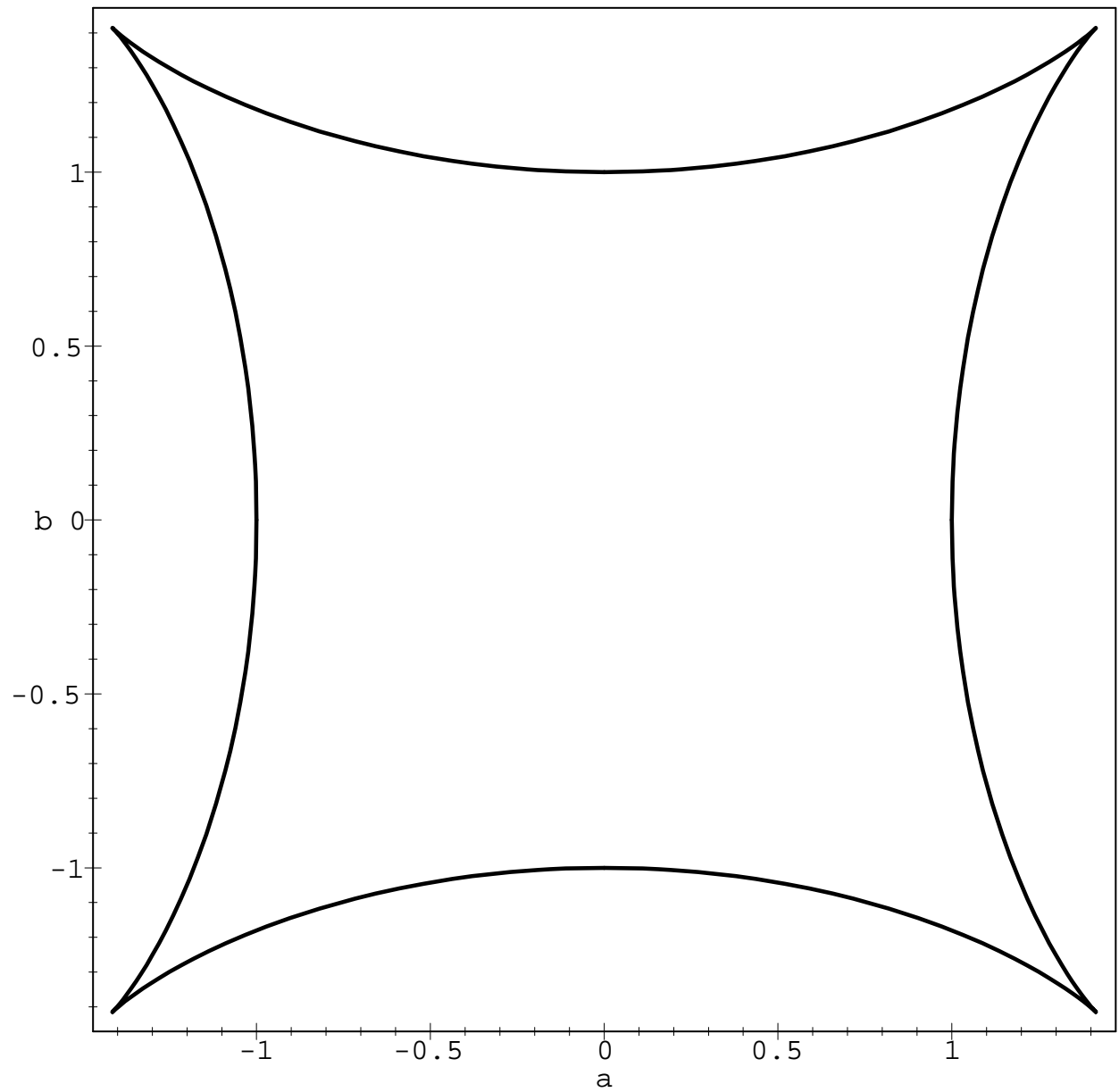
\subsection*{Figure 1. Domain of quasi-Hermiticity at $c=0$}

%


\newpage

\section*{Appendix A: A concise review of the
origin of the present schematic model
 }

The concept of quasi-Hermiticity has been introduced in nuclear
physics \cite{Geyer} where variational calculations of complicated
nuclei proved facilitated by the replacement of the common inner
product $ \langle \psi|\,\phi \rangle$ in Hilbert space by its
generalization
 \be
  \langle \psi|\,
 \Theta\,|\,\phi \rangle
 \,,\ \ \ \ \ \ \Theta=\Theta^\dagger>0\,.
 \label{newpr}
 \ee
Indeed, quantum mechanics can be formulated using {\em any}
invertible and positive definite metric operator $\Theta$ in
(\ref{newpr}). One can feel free to choose any nonstandard
$\Theta\neq I$ and to select observables (i.e., Hamiltonians $H$
etc) represented by operators which are Hermitian with respect to
the new product (\ref{newpr}). Whenever $\Theta\neq I$ we may call
such observables {\em quasi-Hermitian}. In the language of algebra
this means
 \be
 H=H^\ddagger\ \equiv\ \Theta^{-1}\,H^\dagger\,\Theta \,.
 \label{newconj}
 \ee
This condition may be compatible with the manifest non-Hermiticity
of $H$, provided only that the spectrum remains real.
Incidentally, such a reality condition has been found satisfied by
the quartic anharmonic oscillator ``with wrong sign" \cite{BG} (in
this case the coordinate ceases to be observable \cite{Jones}) as
well as by the ``wrong-coupling" cubic oscillator \cite{Caliceti}
(in this model the non-observability concerns its purely imaginary
potential \cite{postc}). Still, a real boom of interest in the
manifestly non-Hermitian quantum Hamiltonians with real spectra
has only been inspired by the well written letter by Bender and
Boettcher in 1998 \cite{BB}. Very persuasive numerical and WKB
arguments have been given there supporting the reality of spectrum
for a broad class of Hamiltonians, with a remarkable impact on
field theory \cite{BBJ}.

The latter class involves the manifestly non-Hermitian
one-dimensional models
 \be
 H = -\frac{d^2}{dx^2} + U(x) + {\rm i}\,W(x) \neq H^\dagger\,
  \label{ptsp}
 \ee
defined on $I\!\!L_2(I\!\!R)$ and containing the two real
components of the potential exhibiting the property  called ${\cal
PT}-$symmetry \cite{BG},
 \be
 {\cal P}\,U(x)\,{\cal P}
 \ (\ \equiv\ U(-x)\ )
 =
 +U(x)\,,\ \ \ \ \ \
  {\cal P}\,W(x)\,{\cal P}=
 -W(x)\,.
 \ee
The rigorous proofs \cite{DDT} of the reality of the spectra (or,
in the present language, of the quasi-Hermiticity) of many
non-Hermitian toy Hamiltonians $H \neq H^\dagger$ proved
complicated but the inconvenience has been circumvented by the
turn of attention to exactly solvable ${\cal PT}-$symmetric
potentials~\cite{solvable}. Their use simplified the proofs and
mathematics but still, our understanding of the ``correct"
assignment of the physical interpretation to a given model
remained incomplete \cite{reviews}. Another simplifying reduction
of the problem was needed and finite-dimensional matrix
Hamiltonians entered the scene \cite{Wang}. In particular, our
recent discussion of the ambiguity of $\Theta$ \cite{Hendrik}
proved best illustrated by the replacement of both the ${\cal
PT}-$symmetric Hamiltonian (\ref{ptsp}) and the operator of parity
${\cal P}$ by the mere two-dimensional, highly schematic matrices
with real elements.

\section*{Appendix B: An irrelevance of the
 shift $c$  }

Even though we confirmed the fourfold symmetry of  $\p {\cal
D}(H^{(3)})$ in paragraph \ref{tasmhle}, this symmetry must be
interpreted as a mere artifact attributed to our choice of the
most comfortable specific $c=0$. Numerical experiments indicate
that the curve $\p {\cal D}(H^{(3)})$ gets deformed and distorted
in proportion to the degree of violation of the equidistance of
the diagonal matrix elements in $H^{(3)}$. Moreover, as long as
all the $c\neq 0$ models (\ref{sekoust}) are characterized by a
not too much more complicated secular equation
 \be
 -{{{E}}}^{3}+\left (3+c\right ){{{E}}}^{2}+\left (
 1-{a}^{2}-{b}^{2}\right ){{E}}-3+3\,{a}^{2}
 -c+c{a}^{2}-{b}^{2}=0\,
 \label{sekous}
 \ee
it would still be feasible to quantify the effect non-numerically.
The method employed in paragraph \ref{tripll} remains applicable
and it leads to an elementary shift of the DEP energy, $3z=3+c$.
The parallel analytic analysis of the $c \neq 0$ problem is left
to the reader. It remains straightforward though a bit boring. For
example, in the most interesting triple-confluence regime the mere
slightly more complicated pair of equations
 \ben
 -3=1-{a}^{2}-{b}^{2}, \ \ \ \ \ \ 1 =-3+3\,{a}^{2}
 -c+c{a}^{2}-{b}^{2}\,
 \een
gives the mere rescaled formulae which relate the DEP matrix
elements in~$H^{(3)}$,
 $$
 a^2=2-\frac{c}{4+c} = 4-b^2. $$
Obviously, the fourfold symmetry of Figure~1 will be broken in a
way which is continuous in the limit of $c \to 0$.

\end{document}